\documentclass[12pt]{iopart}
\usepackage{epsf,psfig}

\begin{document}

\title{Strange Baryon Production From Nucleus-Nucleus Collisions at the AGS, SPS and RHIC}

\author{Huan Z. Huang\dag}

\address{Department of Physics and Astronomy, University of California, Los Angeles, CA 90095-154705, USA}

\address{\dag\ Email: huang@physics.ucla.edu}

\begin{abstract}

We will present a review of selected results on strange baryon production
from nucleus-nucleus collisions at the AGS, SPS and RHIC. 
From the AGS and SPS heavy ion physics program several intriguing
aspects: centrality dependence of strangeness production, anti-hyperon
yield, hyperon production enhancement and baryon number transport 
will be highlighted. Our discussion on RHIC
results will focus on experimental probes of partonic degree of freedom.
Measurements of the elliptic flow $v_{2}$ and the nuclear modification factor 
and their distinctive particle dependence are presented as a function of
transverse momentum.

\end{abstract}

\section{Introduction}

Quantum ChromoDynamics (QCD) calculations on lattice predict the existence
of the Quark Gluon Plasma (QGP), bulk matter of deconfined quarks and gluons~\cite{Karsch}.
The major scientific goal of the heavy ion collision physics is to create
the QGP in high energy nucleus-nucleus collisions and to study QCD
properties of the dense partonic matter. Strangeness, in particular
strange baryon production, plays an important role both as a possible
signature for the QGP formation and as a diagnostic probe of properties of
the partonic matter. We will review experimental data on strange baryon
production in nucleus-nucleus collisions within the context of studying
production dynamics and bulk matter properties.

The production dynamics of strange baryons can reflect the nature of the
partonic matter. It has been argued that gluon fusion processes are
effective channels for strange quark pair production in the QGP leading to
a rapid flavor equilibration such that the production of strange baryons can
be significantly enhanced~\cite{rafelski}. Strangeness equilibrium in a hadronic gas
requires a much longer lifetime for the system; and this is dynamically unfavored
in explosive high energy nucleus-nucleus collisions~\cite{koch}. Finite net baryon
density in nucleus-nucleus collisions at the AGS and SPS will also
impact the strange baryon production dynamics. 
The elliptic flow $v_2$ and the nuclear modification factor will be used to probe 
properties of the matter at early stages of the nucleus-nucleuscollisions.
We will illustrate possible
physical scenarios and production dynamics using experimental data from the
AGS, SPS and RHIC. The selection of the experimental data was based purely on
convenient access of the data and was not meant to be comprehensive. We
apologize to these collaborations whose data have been essential in deepening
our understanding of the baryon production dynamics but are not explicitly
highlighted in this paper.

We conclude this paper by listing several future measurements
which will further elucidate the properties and particle formation dynamics
for the bulk partonic matter.

\section{Selected Topics at the AGS and SPS}

In this section I will briefly discuss several unique physics themes from nucleus-nucleus collisions
at the AGS and SPS and highlight the link between
possible physical scenarios and experimental measurements.

Nucleus-nucleus collisions at the AGS and SPS are characterized by the
formation of high net baryon density. Both the RQMD~\cite{RQMD} and ARC 
model~\cite{ARC} have indicated that the initial net baryon density at the center of central Au+Au
collisions at the AGS is very large, approximately ten times the normal
nuclear matter density. 
One of the conceptual questions arising from such a large baryon density 
is the validity of description in theoretical calculations for hadron scatterings, since hadrons cannot
exist as an independent entity in such a dense environment. The large baryon
density at the initial stage also dictates the evolution dynamics of the
collision: resonances, baryonic resonances in particular, are believed to
play an important role for particle production and collective expansion in
nucleus-nucleus collisions.

The large baryon density has a strong impact on the characterization of
strangeness production in nuclear collisions. Strange and anti-strange
quarks are produced pairwise in nuclear collisions. The strange quarks could
eventually hadronize mostly through associate production ($\Lambda K^{+}$) and kaon
pair production ($K^{+}K^{-}$). At high net baryon densities the Lambda 
($\Lambda $) in the associate production carries the baryon quantum number
from the colliding nuclei; and the energy threshold for the associate
production in N+N$\rightarrow $N$\Lambda K^{+}$ is lower than that for the
kaon pair production in N+N$\rightarrow $NN$K^{+}K^{-}$. The significant
yield of $K^{+}$ from associate production would lead to a large ratio of 
$K^{+}/K^{-}$. Large $K^{+}/K^{-}$ ratios at mid-rapidity from Si+Si, Si+Au and Au+Au
collisions at 14.6, 14.6, 11.7 AGeV incident beam energies, respectively, have been measured
by the E802 collaboration~\cite{e802}. The measured $K^{+}/K^{-}$ ratio
varies from 5 to 7 with little dependence as a function of the number of
participant nucleons. The variation in the ratio is largely due to the beam
energy difference. A large number of $\Lambda$ hyperons, presumably from the associate
production, has also been measured~\cite{e896}.

\begin{figure}[htb]
   \centering
   \epsfxsize=6cm
   \epsfysize=5cm
   \leavevmode
   \epsffile{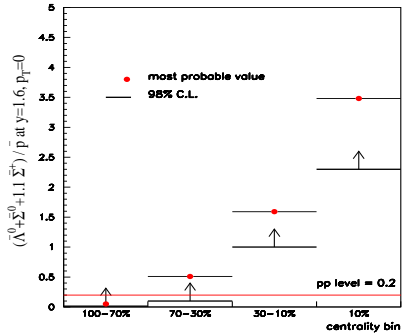} 
\caption[fig:hratio]{Ratio of antihyperon to antiproton as a function of collision 
centralities for Au+Pb collisions at 11.5 A GeV incident energy. Arrows indicate the lower limit of the ratio 
at a $98\%$ confidence level.}
\label{fig:hratio}
\end{figure}

The intricacy of the dynamics in the high baryon density regime is
reflected in the anti-Lambda ($\overline{\Lambda }$) to anti-proton 
($\overline{p}$) ratios. Figure~\ref{fig:hratio} shows the $\overline{\Lambda }$ to 
$\overline{p}$ ratios at $p_{T}$$\sim$$0$ for various centrality bins,
which were derived from E864 and E878 $\overline{p}$ measurements~\cite{e864}. The 
$\overline{\Lambda }$ to $\overline{p}$ ratio is significantly above unity
for the most central Au+Pb collisions at 11.5 A GeV incident energy. 
A ratio greater than one was also obtained by the
E917 experiment~\cite{e917}. The preferred $\overline{\Lambda }$ production 
over $\overline{p}$ could result from dynamics in a
quark clustering scenario: a $\overline{ud}$ diquark is more
likely to find another $\overline{s}$ quark to form a $\overline{\Lambda }$
than to find another $\overline{u}$ quark to form a $\overline{p}$ in
nuclear collisions with high net baryon density where there is an excess of
up and down quarks. The thermal statistical model of the QGP hadronization
essentially depicts this scenario~\cite{rafelski-2}. 
During the hadronic evolution the larger value for $\overline{p}$
annihilations~\cite{stoecker} compared to $\overline{\Lambda }$
annihilations could also lead to a large $\overline{\Lambda }$ to 
$\overline{p}$ ratio in the final state.

\begin{figure}[htb]
   \centering
   \epsfxsize=8cm
   \epsfysize=6cm
   \leavevmode
   \epsffile{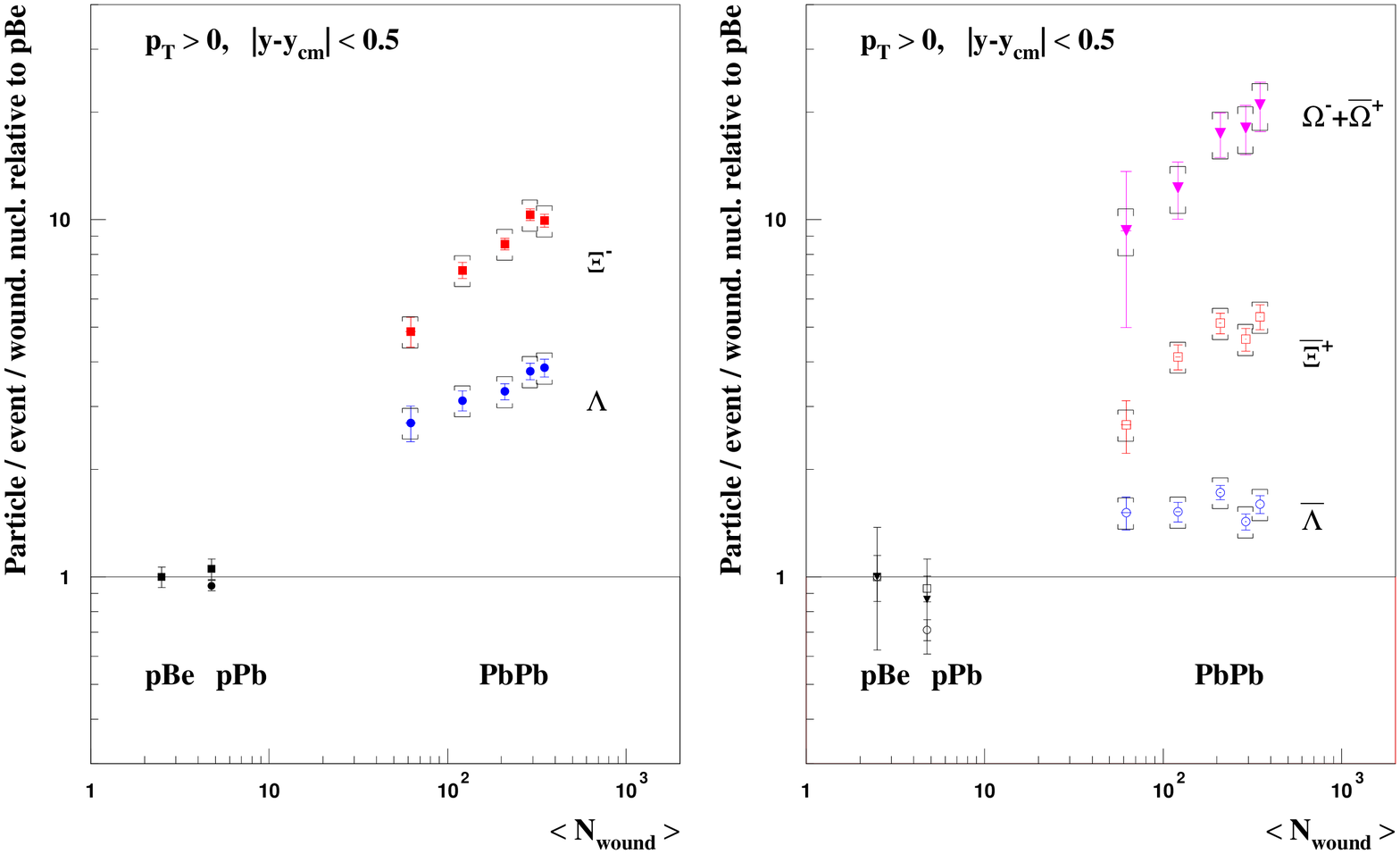} 
\caption[fig:wa97]{Mid-rapidity strange baryon yield per participant nucleon normalized to p+Be collisions measured 
by NA57 at the SPS. The enhancement
factor increases rapidly with the increasing strangeness content from $\Lambda$ to $\Omega$.}
\label{fig:wa97}
\end{figure}

The production of strange baryons ($\Lambda $, $\Xi $ and $\Omega $) and
their anti-particles has been observed to be strongly enhanced with respect
to participant scaling at the SPS energies. Figure~\ref{fig:wa97} presents the mid-rapidity
strange baryon yield per participant (wounded nucleons), normalized to p+Be collisions, measured by the
WA97/NA57 collaboration at the CERN SPS~\cite{WA97}. Similar enhancement factors have also been
reported by the NA49 experiment~\cite{NA49}. The hyperon production per
participant nucleon increases rapidly as a function of the number of
participants. The magnitude of the enhancement increases with the
strangeness content of the hyperons reaching approximately a factor of 20
for ($\Omega +\overline{\Omega }$) in central Pb+Pb collisions at 158 AGeV
incident beam energy.

The magnitude of the enhancement for ($\Lambda$, $\Xi$) is much larger than
that for ($\overline{\Lambda }$, $\overline{\Xi}$). Such an asymmetry between
hyperons and anti-hyperons must be related to the finite net baryon density
at mid-rapidity in nuclear collisions at the SPS. The fragmentation of
colliding nuclei into hyperons is one possible process for the
rapid increase of hyperon yield as suggested by E910~\cite{E910}. 
However, the significant enhancement of anti-hyperon production per
participant indicates that the baryon pair production probability also
increases in central collisions.

\begin{figure}[htb]
   \centering
   \epsfxsize=6cm
   \epsfysize=5cm
   \leavevmode
   \epsffile{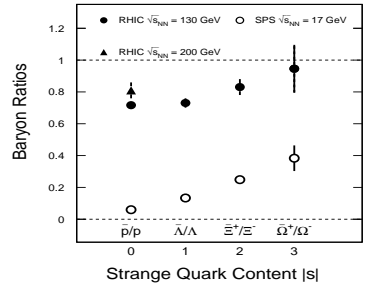} 
\caption[fig:bratio]{Anti-baryon to baryon ratios from central Pb+Pb (SPS) and Au+Au (RHIC) collisions 
as a function of the strangeness content of baryons.}
\label{fig:bratio}
\end{figure}

The fragmentation of the colliding nuclei also contributes to the baryon
number transport in nuclear collisions~\cite{huang}. 
The anti-baryon to baryon ratio from a system with finite net baryon density would be
deviate from unity. Figure~\ref{fig:bratio} shows the anti-baryon to baryon ratio as a
function of the strangeness content of baryons from central Pb+Pb (SPS) and Au+Au (RHIC) collisions. 
The ratios are
much closer to unity at the RHIC energies indicative of a much smaller net
baryon density.

A comment on the $\overline{\Omega }$ to $\Omega $ ratio at the SPS energy
is in order. The ratio is approximately 0.4, significantly different from
unity. This indicates that at mid-rapidity there are 2.5 times more $\Omega $
than $\overline{\Omega }$. Such a large asymmetry probably reflects
intriguing underlying dynamics because strange quarks in $\Omega$($sss$) 
and $\overline{\Omega}$($\overline{s}\overline{s}\overline{s}$) 
must be produced in
pairs in nuclear collisions. The excess of baryon number in $\Omega $ over 
$\overline{\Omega }$ at mid-rapidity is unlikely to be balanced by 
$\overline{\Omega }$ at the forward and backward rapidity regions, since this
would lead to a much wider rapidity distribution 
for $\overline{\Omega }$. In fact, the NA49 experiment 
reported a rapidity width for $\overline{\Omega}$ slightly narrower than that 
for $\Omega$~\cite{NA49}.
The excess of 
$\Omega $ hyperons is likely to carry the baryon numbers from the colliding nuclei.
The dynamical process for baryon number transport from incoming protons and
neutrons to $\Omega $ hyperons which share no up and down valence quarks, is
a subject of intensive theoretical and experimental investigations. Possible
scenarios include direct baryon number transport through gluon junction
interaction and indirect transport through baryon pair production.

\begin{figure}[htb]
   \centering
   \epsfxsize=8cm
   \epsfysize=2.5cm
   \leavevmode
   \epsffile{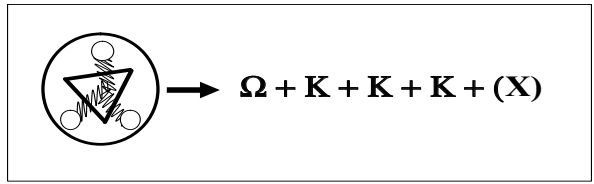} 
\caption[fig:gluon]{Schematic illustration of a baryon in gluon junction configuration. An $\Omega$ hyperon can carry the original 
baryon quantum number if the $\Omega$ is produced during gluon junction interaction where three pairs of $s$-$\overline{s}$ quarks are
created from the gluon string break-ups. Multiple kaons must also be produced in the process.}
\label{fig:gluon}
\end{figure}

Figure~\ref{fig:gluon} depicts a schematic diagram for baryon number transport through
gluon junction interaction. Under strong coupling limit the gluon strings
which are used to model the interactions among valence quarks take the
junction configuration~\cite{khar}. In nuclear collisions if all three gluon strings
are broken up, then the baryon quantum number is transported by the gluon
junction. The probability for such gluon junction interactions could be
significantly enhanced in nucleus-nucleus collisions leading to a larger
degree of baryon stopping at mid-rapidity~\cite{vance}. For the $\Omega $ to be produced
from the gluon junction interaction and carry the baryon number from colliding nuclei
multiple kaons must also be produced. The correlation of these kaons with the 
$\Omega $ provides a unique experimental signature for this process.

Hyperon pair production can proceed through channels of 
$\Omega$-$\overline{\Omega }$ or $\Omega $-$\overline{\Xi }K$, $\Xi $-$\overline{\Xi}$ 
or $\Xi $-$\overline{\Lambda }K$, and $\Lambda $-$\overline{\Lambda }$ or 
$\Lambda $-$\overline{p}K$. The kaon associated channels are normally
suppressed compared to the other due to a higher energy threshold. However, the
presence of finite net baryon density could modify the relative fraction of
these channels. The $\overline{\Omega }$ to $\Omega $ ratio of 0.4 would
indicate that the $\Omega $-$\overline{\Xi }K$ channel is 
enhanced over the $\Omega $-$\overline{\Omega }$ channel for $\Omega $
production at mid-rapidity if
these channels dominate the $\Omega $ production. Measurements
of correlations among $\Omega $, $\overline{\Xi }$ and kaons can shed light
on the production dynamics.

\section{Strange Baryon Production at RHIC}

The yield of $\Lambda $ and $\overline{\Lambda }$ production at RHIC was
first reported by STAR~\cite{starL} and PHENIX~\cite{phenixL} for Au+Au collisions at 
$\sqrt{s_{NN}}=130$ GeV. The measured rapidity density ($dn/dy$) of $\Lambda $ and 
$\overline{\Lambda }$ depends linearly on the pseudo-rapidity density of
negatively charged hadrons. The linear dependence of the
baryon rapidity density on charged hadrons (mostly pions) is also a feature
in baryon production models based on string fragmentations. However, in
fragmentation models such as HIJING~\cite{xnwang} the yields of strange hyperons are
greatly suppressed: the HIJING model produces too many anti-protons and too few 
$\Lambda $ and $\overline{\Lambda }$ hyperons. The discrepancy between data
and the fragmentation model calculation is even larger for multi-strange
hyperons. By introducing final state interactions between co-moving hadrons
or partons from densely populated interacting strings, the yield of
multiple-strange hyperons can be significantly enhanced from primordial
string fragmentation processes~\cite{Capella}. The most effective reactions are 
$\pi N\rightarrow K\Lambda (\Sigma )$, $\pi \Lambda (\Sigma )\rightarrow K\Xi$, 
$\pi \Xi \rightarrow K\Omega $ and their respective anti-particle
reactions.

In addition, multi-parton dynamics such as
gluon junction hadronization~\cite{khar,vance}, quark coalescence~\cite{lin,molnar,greco} and parton
recombination~\cite{hwa,fries} have been applied to strange baryon production.
Since based on parton structure
functions of nuclei gluons are expected to be mainly responsible for the
formation of the initial high energy density matter in nucleus-nucleus
collisions at RHIC, the mechanism of gluon junction hadronization for baryon
production is an intriguing scenario.
With the formation of a high density gluonic fireball,
the gluon junction depicted in Figure~\ref{fig:gluon} would naturally be a possible
topological configuration existing inside the fireball. Then the
simultaneous breakups of the three gluon strings of the junction would lead
to the formation of a baryon or an anti-baryon. In such a formation scheme
the mass of the baryon could come mainly from the gluon junction and the
flavor of the baryon would be determined by the light quark pair from the
gluon string breakup. As a result the production probability of multiple
strange hyperons would not be significantly suppressed. 
In the normal string fragmentation mechanism for baryon production hyperons
are strongly suppressed due to small tunneling probability for the higher mass
diquarks of the hyperons~\cite{string,lund}. In quark coalescence or quark recombination
scenarios, if the initial quark (e.g., constituent quarks in ALCOR model~\cite{alcor})
matter has sufficiently large number of strange quarks available, the
production of high mass hyperons can also be enhanced. In all these
scenarios the formation of baryons requires the interaction of multiple
gluons or quarks (partons) in contrast to standard single parton
fragmentation where the rest of the baryon constituents being treated as out
of vacuum.

\begin{figure}[htb]
   \centering
   \epsfxsize=8cm
   \epsfysize=6cm
   \leavevmode
   \epsffile{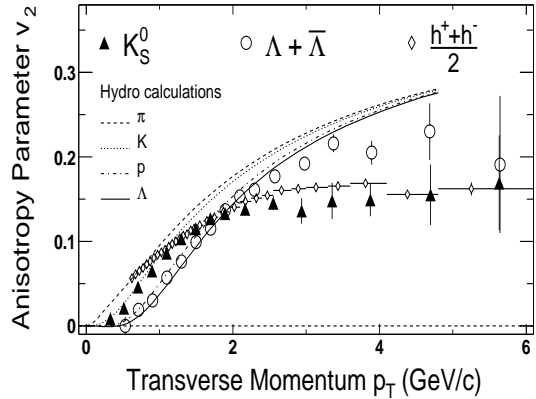} 
\caption[fig:star_v2]{Elliptic flow $v_2$ as a function of transverse momentum for $K_S$,
 $\Lambda +\overline{\Lambda }$ and charged hadrons from minimum bias Au+Au collisions
at $\sqrt{s_{_{NN}}}$ 200 GeV. Predictions of hydrodynamical calculations for identified particles
are shown for comparison.}
\label{fig:star_v2}
\end{figure}

Identified particles, especially with comparisons between baryons and
mesons, provide a unique means to investigate possible formation dynamics
from the partonic perspective and to examine properties of the partonic matter
carried by hadrons in the final state~\cite{sorensen,long,julia}. 
Figure~\ref{fig:star_v2} shows $v_{2}$ as a function of transverse momentum $p_{T}$ for 
$K_{S}$, $\Lambda +\overline{\Lambda }$ and charged hadrons from minimum bias Au+Au collisions
at $\sqrt{s_{_{NN}}}$ 200 GeV. The $v_{2}$ of heavy particles 
$\Lambda + \overline{\Lambda }$ is smaller than that of light $K_{S}$ 
in the low $p_{T}$ region below 1.5 GeV/c. This $v_{2}$ ordering of
particle dependence is consistent with hydrodynamical calculations where
parton thermalization and collective expansion velocity driven by a pressure
gradient are assumed. Note that inclusive transverse momentum distributions of identified
pions, kaons and protons at low $p_{T}$ can also be described in this
framework, for example using a blast-wave parameterization for collective
motions~\cite{blfit}.

At the intermediate $p_{T}$ above 2 GeV/c, however, the $v_{2}$ of hyperons
is larger than that of $K_{S}$ and the magnitude of $v_{2}$ shows little 
$p_{T}$ dependence. Both features are in contradiction with hydrodynamical
calculations~\cite{hydro}. A $p_{T}$ independent $v_{2}$ may result from particle
emission from the surface of an ellipsoid formed in nucleus-nucleus
collisions~\cite{shuryak}. A surface emission scenario can be accommodated dynamically if
at the time of the particle emission the source is at sufficiently high
density and opaque. If the particle formation is through the fragmentation
of high $p_{T}$ partons which have undergone partonic energy loss in the
medium, the large $v_{2}$ of hyperons would imply a larger energy loss for
partons of hyperon production than those of kaons.

\begin{figure}[htb]
   \centering
   \epsfxsize=10cm
   \epsfysize=8cm
   \leavevmode
   \epsffile{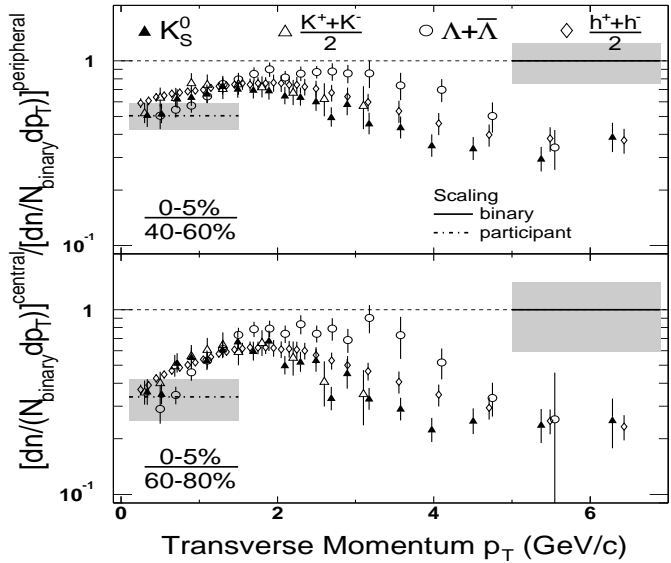} 
\caption[fig:star_raa]{Nuclear modification factor $R_{CP}$ as a function of transverse momentum for charged kaons, 
$K_S$, $\Lambda + \overline{\Lambda }$ and charged hadrons. The solid (dashed) lines show the values expected for binary collision
(participant) scaling where the widths of the bands indicate the systematic errors from the $N_{binary}$ and $N_{part}$ 
calculations.}
\label{fig:star_raa}
\end{figure}

Figure~\ref{fig:star_raa} shows the nuclear modification factor ($R_{CP}$) derived from ratios for
particle yields of central to peripheral collisions normalized to the number
of binary collisions, for hyperons and kaons, where 
\begin{equation*}
R_{CP}=\frac{[d^{2}n/(N_{binary}dp_{T}dy)]^{central}}{[d^{2}n/(N_{binary}dp_{T}dy)]^{peripheral}}.
\end{equation*}
Details of the measurement can be found in references~\cite{long,norman,star_pt,star_prl}. The
suppression of particles at the intermediate $p_{T}$ region with respect to
the binary collision scaling is much smaller for hyperons. In fact, there
is an absence of suppression for hyperons for $p_{T}$ from 1.8 to 3.5 GeV/c
whereas the kaons are suppressed over all 
$p_{T}$. If the nuclear modification is dominated by the process of partonic
energy loss followed by parton fragmentation, then the smaller suppression
of the hyperons would imply a smaller energy loss, contradicting the
larger energy loss needed for a greater $v_{2}$.

Distinctive $p_{T}$ scales associated with
different dynamical processes in nucleus-nucleus collisions at RHIC have been observed: parton
fragmentation at $p_{T}$ greater than 6 GeV/c where a suppression of particle yield with 
respect to binary collision scaling has been observed~\cite{phenix_hipt,star_highpt}; and hydrodynamical behavior
at $p_{T}$ less than 2 GeV/c. The intermediate $p_{T}$ region of 2 to 6
GeV/c is uniquely sensitive to multi-parton dynamics, such as gluon junction
hadronization for baryons~\cite{vance}, quark coalescence~\cite{molnar,greco} and parton 
recombinations~\cite{hwa,fries}. That sensitivity may be reflected in the elliptic flow $v_{2}$ and nuclear
modification factor measurement. In a quark coalescence or
recombination scenario the elliptic flow of a particle results from an
anisotropical angular distribution of individual quarks/partons at the time of
hadron formation; and the nuclear modification factor measures the
dependence of particle formation probability on the parton density. In these
scenarios it is natural to expect the baryon formation probability to depend
on higher powers of parton density than mesons. Therefore, the yield of
baryons in the intermediate $p_{T}$ region increases faster than that of
mesons from peripheral to central collisions. Recent STAR~\cite{star_prl} and 
PHENIX~\cite{phenix_prl}
measurements reveal qualitatively an important contribution of multi-parton
dynamics in the intermediate $p_{T}$ region. To establish the exact
dynamical scenario requires more experimental measurements on a variety of
hyperons and mesons and deeper theoretical understanding of the
hadronization phenomenology.

\section{Perspective}

RHIC as a dedicated nuclear QCD research facility at high energy has
provided a unique opportunity to study partonic dynamics in nucleus-nucleus
collisions. Experimental data from RHIC have shown prominent features
consistent with collective partonic dynamics, which are absent or overwhelmed by
hadronic features in low energy data. In nucleus-nucleus collisions at RHIC
we can identify three distinct $p_{T}$ scales: the hydrodynamic region
below 2 GeV/c where properties of bulk matter due to interactions at the
partonic phase or the hadronic phase or both are important; the fragmentation
region above 6 GeV/c where particle production through parton fragmentation
may become dominant; and the intermediate $p_{T}$ region of 2 to 6 GeV/c
where multi-parton dynamics may be prominent with enhanced sensitivity to
partonic properties of the bulk matter. We will list several future
measurements which will shed light on both the transition among these $p_{T}$
regions and the underlying dynamics of partonic nature.

The STAR $v_{2}$ and $R_{CP}$ measurements of $\Lambda +\overline{\Lambda }$
and $K_{S}$ along with charged hadrons indicate that the particle
dependence of the nuclear modification factor disappears for $p_{T}$ greater
than 5-6 GeV/c. The jet quenching phenomenon~\cite{xnwang}, if it is proven to be applicable
at the high $p_{T}$ region, will provide a unique probe of the dense
matter at the initial stage of nuclear collisions. The angular anisotropy 
$v_{2}$ for identified particles in this high $p_{T}$ region is an important
measurement to establish the validity of the pQCD-inspired fragmentation
approach for particle production. The combination of $v_{2}$ and nuclear modification factor
measurements for identified particles above $p_{T}$ 6 GeV/c can provide
quantitative constraints on possible energy loss of partons through a dense hot QCD medium.

In order to verify the possible geometrical origin of the $p_{T}$ independent
saturated  $v_{2}$ value, it is important to measure elliptic flow at
intermediate $p_{T}$ from light ion collisions. In collisions of light ions
the surface to volume ratio should be very different from Au+Au collisions
which may lead to a different $p_{T}$ dependence for $v_{2}$. The nuclear
modification factor in light ion collisions may also be different and
systematic studies of A dependence may lead to better 
quantitative measures related to the physical
origin of the high $p_{T}$ suppression.

Particles with small hadronic rescattering cross sections suffer small final
hadronic interactions with co-moving hadrons and therefore may carry the
information of matter properties at the particle formation time. It is
generally believed that $\phi $ meson, $\Omega $ ($\Xi $ and $\Lambda $) and
heavy quark particles (mostly $D$ mesons at RHIC) satisfy the requirement
for small hadronic cross sections and can be used to probe partonic matter
properties~\cite{xnu}. Note that the information derived from these particles provides
a snapshot of the matter properties at the particle formation time, the so
called chemical freeze-out time, which lies at the phase boundary between
quarks and hadrons. Such snapshot information complements that from the
electromagnetic probes such as leptons which provide a time-integrated evolution
history of the dense matter. A vigorous program to measure $v_{2}$ and the
nuclear modification factor for $\phi $, $\Omega $ and $D$ particles is
being pursued.

Identified particle correlations over a large $\Delta y$-$\Delta p_{T}$
scale such as those among $\Omega $-kaons and $\Omega $-$\overline{\Xi }$
will shed important insight on possible roles of multi-parton dynamics in
nucleus-nucleus collisions at RHIC. The gluon junction interaction and its
possible role in baryon number transport and baryon production remains an
intriguing theoretical suggestion. In
order to carry out these correlation measurements at RHIC with STAR, several
upgrades are required such as the barrel Time-of-Flight detector, the micro-Vertex
detector and improvements in the TPC front-end electronics and data acquisition system. 

\section{Acknowledgment}

We thank Hui Long, Johann Rafelski, Hans Georg Ritter, Raimond Snellings,
Paul Sorensen, An Tai, Nu Xu and Zhangbu Xu for enlightening discussions on
topics covered in this review. Charles Whitten carefully read and improved
the manuscript. We thank the SQM03 organizers for the invitation to present
this review and for the successful conference at North Carolina.


\section*{References}

\begin{thebibliography}{99}
\bibitem{Karsch} Karsch F. 2002 Nucl. Phys. A{\bf 698} 199c
\bibitem{rafelski} Rafelski J and Muller B 1982 Phys. Rev. Lett. {\bf 48} 1066
\bibitem{koch} Koch P, Muller B and Rafelski J 1986 Phys. Rept. {\bf 142} 167
\bibitem{RQMD} Sorge H {\it et al} 1989 Nucl. Phys. A{\bf 498} 567c
\bibitem{ARC} Kahana S H {\it et al} 1996 Ann. Rev. Nucl. Part. Sci. {\bf 46} 31
\bibitem{e802} Abbott T {\it et al} 1990 Phys. Rev. Lett. {\bf 64}  847; Ahle L {\it et al} 1999
	Phys. Rev. C{\bf 60} 044904
\bibitem{e896} Albergo S {\it et al} 2002 Phys. Rev. Lett. {\bf 88} 062301
\bibitem{e864} Armstrong T {\it et al} 1999 Phys. Rev. C{\bf 59} 2699; Bennett M {\it et al} 1997
 	Phys. Rev. C{\bf 56} 1521
\bibitem{e917} Back B B {\it et al} 2001 Phys. Rev. Lett. {\bf 87} 242301
\bibitem{stoecker} Stoecker H {\it et al} 1995 Nucl. Phys. A{\bf 590} 271
\bibitem{rafelski-2} Rafelski J and  Letessier J 1999 Acta Phys. Polon. B{\bf 30} 3559
\bibitem{WA97} Antinori F. {\it et al} 2002 J. Phys. G: Nucl. Part. Phys. {\bf 28} 1607 and these proceedings
\bibitem{NA49} Afanasiev S V {\it et al} 2002 Preprint nucl-ex/0208014 and these proceedings
\bibitem{E910} Chemakin I {\it et al} 2000 Phys. Rev. Lett. {\bf 85} 4868
\bibitem{huang} Huang HZ 2002 J. Phys. G: Nucl. Part. Phys. {\bf 28} 1667 
\bibitem{khar} Kharzeev D 1996 Phys. Lett. B{\bf 378} 238
\bibitem{vance} Vance S E {\it et al} 1998 Phys. Lett. B{\bf 443} 45
\bibitem{starL} Adler C {\it et al} 2002 Phys. Rev. Lett. {\bf 89} 092301
\bibitem{phenixL} Adcox K {\it et al} 2002 Phys. Rev. Lett. {\bf 89} 092302
\bibitem{xnwang} Wang X N and Gyulassy M 1992 Phys. Rev. Lett. {\bf 68} 1480
\bibitem{Capella} Capella A 2003 Preprint nucl-th/0303045
\bibitem{lin} Lin Z W and Ko C M 2002 Phys. Rev. Lett. {\bf 89} 202302
\bibitem{molnar} Molnar D and Voloshin S A 2003 Preprint nucl-th/0302014
\bibitem{greco} Greco V, Ko C M and Levai P 2003 Preprint nucl-th/0301093
\bibitem{hwa} Hwa R C and Yang C B 2003 Phys. Rev. C {\bf 67} 034902
\bibitem{fries} Fries R J, Muller B, Nonaka C and Bass S A 2003 Preprint nucl-th/0301087
\bibitem{string} Casher A, Neuberger H and Nussinov S 1979 Phys. Rev. D {\bf 20} 179
\bibitem{lund} Bo Andersson ``The Lund Model'' Cambridge University Press
\bibitem{alcor} Biro T S, Levai P and Zimanyi J 1995 Phys. Lett. B {\bf 347} 6; 1999 Phys. Rev. C {\bf 59} 1574
\bibitem{sorensen} Sorensen P these proceedings
\bibitem{long} Long H these proceedings
\bibitem{julia} Velkovska J these proceedings
\bibitem{blfit} Xu N and Kaneta M 2002 Nucl. Phys. A {\bf 698} 306c
\bibitem{hydro} Huovinen P {\it et al} 2001 Phys. Lett. B {\bf 503} 58
\bibitem{shuryak} Shuryak E V 2002 Phys. Rev. C {\bf 66} 027902
\bibitem{norman} Norman B these proceedings
\bibitem{star_pt} Adams J {\it et al} 2003 Preprint nucl-ex/0305015
\bibitem{star_prl} Adams J {\it et al} 2003 Preprint nucl-ex/0306007
\bibitem{phenix_hipt} Adcox K {\it et al} 2002 Phys. Rev. Lett. {\bf 88} 022301
\bibitem{star_highpt} Adler C {\it et al} 2002 Phys. Rev. Lett. {\bf 89} 202301
\bibitem{phenix_prl} Adler S S {\it et al} 2003 Preprint nucl-th/0305036
\bibitem{xnu} Xu N 2002 J Physics G: Nucl. Part. Phys. {\bf 28} 2121

\end{thebibliography}
\end{document}